\documentclass[twocolumn,showpacs,preprintnumbers,amsmath,amssymb]{revtex4}

\usepackage{graphicx}
\usepackage{dcolumn}
\usepackage{bm}
\usepackage{amssymb}

\begin{document}

\preprint{APS/123-QED}

\title{Asymptotic evolution of quantum walks on the $N$-cycle subject to decoherence on both the coin and position degrees of freedom}

\author{Chaobin Liu}
 \email{cliu@bowiestate.edu}
\author{Nelson Petulante}%
 \email{npetulante@bowiestate.edu}
\affiliation{%
Department of Mathematics, Bowie State University, Bowie, MD, 20715 USA\\
}%
\date{\today}
\begin{abstract} 

Consider a discrete-time quantum walk on the $N$-cycle subject to decoherence both on the coin and the position degrees of freedom. By examining the evolution of the density matrix of the system, we derive some new conclusions about the asymptotic behavior of the system. When $N$ is odd, the density matrix of the system tends, in the long run, to the maximally mixed state, independent of the initial state. When $N$ is even, although the behavior of the system is not necessarily asymptotically stationary, in this case too an explicit formulation is obtained of the asymptotic dynamics of the system. Moreover, this approach enables us to specify the limiting behavior of the mutual information, viewed as a measure of quantum entanglement between subsystems (coin and walker). In particular, our results provide efficient theoretical confirmation of the findings of previous authors, who arrived at their results through extensive numerical simulations. Our results can be attributed to an important theorem which, for a generalized random unitary operation, explicitly identifies the structure of all of its eigenspaces corresponding to eigenvalues of unit modulus.

\end{abstract}

\maketitle

\section{INTRODUCTION}

In recent times, quantum walks (QW) have attracted extensive attention, mainly for their value as potential sources of new algorithms \cite{CG04, A07,VA08, PGKJ09}. However, quantum walks are physically and mathematically interesting in their own right due to the richness of dynamical and statistical properties inherent in the systems \cite{K03, K08, GJS04}.

In stride with the theoretical advances, several schemes, such as \cite{KFC09, SM10, ZKG10, ZLLLLG10, SCP10, BFL10},  have been proposed to implement quantum walks in realistic media. However, any attempt to implement a quantum system in a physical channel must take into consideration the critical issue of ``decoherence". As the aptly contrived neologism suggests, decoherence connotes disruption of the characteristic coherent features of a quantum system, including entanglement of subsystems. Subject to decoherence, a quantum system tends to behave, in the long-term limit, like a classical system.




Various mathematical models of decoherence in discrete-time QWs have been proposed and investigated both numerically and analytically  \cite{BCA022, KT032, RSAAD05, R07, BSCR08, LP092, AAAD09, SCSK10}. In general, regardless of the model adopted, a discrete-time QW, when subject to decoherence, tends to revert, in the long-term limit,  to its classical analogue. However, when properly understood and appropriately controlled, it turns out that decoherence actually can serve to improve the algorithmic properites of a quantum walk \cite{K06}.

In this paper, we investigate the evolution of a quantum walk on the $N$-cycle under the assumption of decoherence-inducing disturbances on both the coin and the position degrees of freedom. As evidenced by the literature, much has been written about the case of a purely coherent QW on the $N$-cycle. For instance, in \cite{AAKV01} it is shown that if the parity of the cycle-length $N$ is odd, then the time-averaged distribution of a coin-governed quantum walk on the $N$-cycle mixes to a uniform distribution. On the other hand, according to \cite{TFMK03, BGKLW03}, if the parity of $N$ is even, then the limiting distribution still exists, but might be non-uniform. For an interesting investigation of the fluctuation of quantum walks on the $N$-cycle, as characterized by their temporal standard deviation, the reader is referred to \cite{IKKS05}.

For the QW on the $N$-cycle, the mathematical model of decoherence adopted in this paper can be described as follows. At every time step of the walk, the option persists, with constant probability $q$, of exercising a projective measurement on the basis states of the quantum system. As such, three distinct cases emerge: decoherence might be assumed to apply to 1) the position only, 2) to the coin only, or 3) jointly to the position and coin.  Analytic treatments of case 1) can be found in \cite{KT032, R07}. For an analytic treatment of case 2), see  \cite{LP092,LP10}. Numerical treatments of all three cases can be found in  \cite{KT032, MK07}. Judging by the literature, analytic treatments of case 3) are rare at best. In this paper, our main objective is to provide a thorough analytic treatment of case 3).

By examining the linear structure of the quantum operation underlying the QW and analyzing the evolution of the associated quantum Markov chain, our approach yields an explicit formulation of limiting density matrix of the quantum system, both for odd and even values of the cycle length $N$. This approach not only allows us to derive the limiting probability distribution, but also enables us to formulate a clear account of the limiting behavior of the system's quantum entanglement features.

In what follows, Section 2 is devoted to introducing and developing the basic ingredients essential to our investigation. In particular, for {\it a generalized random unitary operation}, we specify the linear structure of its key eigenspaces. We proceed, in section 3, to examine the Markov chain evolution of the density operator of a quantum walk on the $N$-cycle subjet to decoherence on both the coin and the position degrees of freedom. Our analysis includes a consideration of the effect of decoherence upon entanglement. In Section 4, we offer some concluding remarks. Finally, the proofs of the theorems and theorem-like assertions are presented in Appendices A through E.

\section{Generalized random unitary operation}

A well-known model of state transition maps for quantum systems is provided by {\em random unitary operations} \cite{ASH01, AS08}. In this section, we introduce a generalization of the random unitary operation model. By appending to the usual definition of a random unitary operation a non-unitary bistochastic component, the resulting model is inclusive enough to lend itself to an analysis of the long-term tendencies both of decoherence and entanglement for a quantum walk on the $N$-cycle.    

We begin with some preliminaries. Given a Hilbert space $\mathcal{H}$ of finite dimension $m$, let $\mathfrak{B}(\mathcal{H})$ denote the set of all linear operators on $\mathcal{H}$ with inner product defined by
\begin{eqnarray}
\langle X,Y \rangle \equiv \mathrm{tr}( X^{\dagger}Y).\label{innerproduct}
\end{eqnarray}

The corresponding norm, called {\em Frobenius norm} or {\em Schatten} 2-{\em norm}, is defined by  
\begin{eqnarray}
\|X\| \equiv [\mathrm{tr}( X^{\dagger}X)]^{1/2}=\sqrt{\langle X,X \rangle}. \label{norm}
\end{eqnarray}
This choice of norm on $\mathfrak{B}(\mathcal{H})$ will remain in effect throughout this paper. 

Let $\mathfrak{D}(\mathcal{H}) \subset \mathfrak{B}(\mathcal{H})$ denote the set of {\em positive} operators $\rho: \mathcal{H}\rightarrow \mathcal{H}$ with Tr$(\rho)=1$. The operators $\rho \in \mathfrak{D}(\mathcal{H})$ are the so-called ``density operators". They serve to model, as faithfully as do the ``state vectors" themselves, the possible states of a quantum system whose state vectors reside in $\mathcal{H}$.  
 
By a {\em super-operator} ${\bf \Phi}$ on $\mathfrak{B}(\mathcal{H})$, we mean a linear mapping ${\bf \Phi}: \mathfrak{B}(\mathcal{H})\rightarrow\mathfrak{B}(\mathcal{H})$, with norm defined by  
\begin{eqnarray}
\|{\bf \Phi}\| \equiv \mathrm{Sup}_{X\in \mathfrak{B}(\mathcal{H})}\frac{\|{\bf \Phi}(X)\|}{\|X\|}.
\end{eqnarray}

Note that $\mbox{dim}{\mathfrak{B}(\mathcal{H})}=m^{2}$, where $m=\mbox{dim}(\mathcal{H})$. Thus, any super-operator ${\bf \Phi}$ on $\mathfrak{B}(\mathcal{H})$ can be represented, relative to a given basis for $\mathfrak{B}(\mathcal{H})$, by an $m^{2}\times m^{2}$ matrix. In the sequel, this matrix will be denoted by the symbol $\left[{\bf \Phi}\right]$. In particular, relative to a special basis consisting of eigenvectors and generalized eigenvectors of ${\bf \Phi}$, the shape of the matrix $\left[{\bf \Phi}\right]$ conforms to a special quasi-diagonal lay-out called the Jordan canonical form. The details can be found in any one of a number of sources, including \cite{NAJ10}.  

Among the set of super-operators, we distinguish a special subset called ``quantum operations". By definition, to qualify as a quantum operation, the super-operator ${\bf \Phi}$ must be {\em completely positive}, meaning that the extended map ${\bf \Phi}\otimes \mathbb{I}_n$ is positive for all $n\geq 1$. 

The formalism of quantum operations is versatile enough to handle both unitary (closed) and non-unitary (open), or a mixture thereof, of discrete transitions of state of a quantum system. For a good introductory exposition of this subject, see \cite{NC00,P08}.

By Choi's Theorem \cite{Choi1975} and \cite{NC00, P08, K1983}, any completely positive linear operator, including any quantum operation ${\bf \Phi}: \mathfrak{B}(\mathcal{H})\rightarrow\mathfrak{B}(\mathcal{H})$, can be represented in terms of a set $\mathcal{A}=\{A_i \,|\,\, i=1,2,..., m^{2}\}$ of ``Kraus operators", as follows:  

\begin{eqnarray}
{\bf \Phi}_{\mathcal{A}}(X)=\sum_{i}A_iX A_i^{\dagger}.\label{choi-ex}
\end{eqnarray}
In this expression, which we call the ``Choi expansion" of ${\bf \Phi}$, the symbol $A_i^{\dagger}$ denotes ${\bar{A_i}}^{T}$(transpose of the complex conjugate of $A_i$). 
 
In terms of the Choi expansion, the condition of being {\em trace-preserving}, meaning that $\mbox{Tr}({\bf \Phi}_{\mathcal{A}}(X))=\mbox{Tr}(X)$ for all $X \in \mathfrak{B}(\mathcal{H})$, is equivalent to the condition:

\begin{equation}
{\sum_{i}A_i^{\dagger}A_i=\mathbb{I}_m}.\label{tracepreserving}  
\end{equation}

On the other hand, if the Kraus operators of ${\bf \Phi}_{\mathcal{A}}$ satisfy the dual condition:
\begin{equation}
{\sum_{i}A_i A_i^{\dagger}=\mathbb{I}_m},\label{unital}
\end{equation} 
then ${\bf \Phi}_{\mathcal{A}}$ is said to be {\em unital}.  
Note that Eq. (\ref{unital}) is equivalent to the simple statement that ${\bf \Phi}_{\mathcal{A}}(\mathbb{I}_m)=\mathbb{I}_m$.



A quantum operation which is both unital and trace-preserving is called {\em bistochastic}. It is a routine matter to verify that any convex linear combination of bistochastic quantum operations is itself a biostochastic quantum operation. 

Now suppose the bistochastic quantum operation ${\bf \Phi}$  admits a convex decomposition of the form
\begin{eqnarray}
{\bf \Phi}(\rho)=\sum_{i=1}^k p_iU_i\rho U_i^{\dagger}+q\sum_{j=1}^l A_j \rho A_j^{\dagger},\label{GRO}
\end{eqnarray}
where the operators $U_{i}$ are unitary, the operators $A_{j}$ satisfy the conditions Eqs. (\ref{tracepreserving}) and (\ref{unital}), and where, of course, $q+\sum_{i=1}^kp_i=1$.
\ For lack of a better term, a bistochastic quantum operation ${\bf \Phi}$ admitting a convex decomposition as in Eq. (\ref{GRO}) is called a {\em generalized random unitary operation} (henceforth abbreviated GRO).

For the purposes of this study, the action of a GRO is interpreted as follows. At each time step of the quantum walk, the decoherence effect is delivered by the summation term in Eq.(\ref{GRO}) involving $\{A_j\}_{j=1}^{l}$, whose coefficient $q$ denotes the {\em decoherence rate}. Meanwhile, at each time step of the walk, the complimentary summation term of Eq.(\ref{GRO}) involving $U_i$ imparts a purely unitary transition of quantum state whereby each of the $U_i$ is applied to the system with corresponding probability $p_i$.   

The GRO model embraces, as special cases, several types of quantum operations prevalent in the literature. For instance, when $q=0$, the GRO model equates to the usual model of random unitary operations as in \cite{NAJ09, NAJ10}. When $U_i$ and $A_j$ are appropriately customized, the resulting GRO equates to the model of decoherence, as in \cite{KT032}, for coin-governed quantum walks. 


Given a bistochastic quantum operation ${\bf \Phi}$, at least one of its eigenvalues $\lambda$ must belong to the unit circle (i.e. $|\lambda|=1$). In particular, this is so if ${\bf \Phi}$ is a GRO as in Eq. (\ref{GRO}). For an eigenvalue $\lambda$ of ${\bf \Phi}$, let $\mathsf{Ker}({\bf \Phi}-\lambda \mathbb{I})$ denote the eigenspace of $\lambda$. Essential to the aims of this investigation is a determination of the structure of eigenspace $\mathsf{Ker}({\bf \Phi}-\lambda\mathbb{I})$. The following theorem represents a significant step in this direction.  

{\bf Theorem 1.}\,\, Let ${\bf \Phi}: \mathfrak{B}(\mathcal{H})\rightarrow \mathfrak{B}(\mathcal{H})$ be the generalized random unitary operation given by 
\begin{equation}
{\bf \Phi}(\rho)=\sum_{i=1}^k p_iU_i\rho U_i^{\dagger}+q\sum_{j=1}^l A_j \rho A_j^{\dagger}, \nonumber
\end{equation}
and let $\lambda$, with $|\lambda|=1$, be an eigenvalue of ${\bf \Phi}$. Then $X\in \mathrm{Ker}({\bf \Phi}-\lambda \mathbb{I})$ if and only if for each index $1\leq i\leq k$ we have $U_iX=\lambda XU_i$ and $U_iXU_i^{\dagger}=\sum_{j=1}^lA_jXA_j^{\dagger}$.


Proof. See Appendix A.\vspace*{5pt}

Theorem 1 generalizes a result in \cite{NAJ10}. Its proof relies on a pattern of reasoning similar to that employed in \cite{NAJ10} and the property of the contractivity of a completely positive trace-preserving linear map \cite{PGWPR06}.

As shown in literature (e.g. \cite{MW10, LPQMC10}), for a bistochastic quantum operation, the eigenvalues lying on the unit circle determine the evolution of the associated quantum Markov process, including the existence or non-existence of a long-term stationary state. More precisely, the long-term behavior of the quantum Markov process is linked intimately to the structure of the eigenspaces of eigenvalues on the unit circle. For an arbitrary quantum operation, the problem of determining explicitly all its eigenvalues on the unit circle remains intractable. However, at least for the specific types of quantum operations, theorem 1 provides an efficient means for identifying the eigenvalues and the eigenspaces of eigenvalues of absolute value 1. The following corollary provides an example of this. Nontrivial applications of theorem 1 are relegated to the next section. 

{\bf Corollary 2.}\,\, Suppose the bistochastic quantum operation ${\bf \Phi}$ is given by 

\begin{eqnarray}
{\bf \Phi}(X)=(1-q)X+q\sum_{i}A_iX A_i^{\dagger},
\end{eqnarray}
where $\sum_{i}A_iA_i^{\dagger}=q\mathbb{I}$ and $0<q<1$. Then $\lambda=1$ is the only eigenvalue on the unit circle.

\section{Quantum walks on the $N$-cycle exposed to decoherence both on coin and position degrees of freedom}

For a quantum walk on the $N$-cycle, the {\em position space} of the walker is the Hilbert space $\mathcal{H}_{N}$ spanned by an orthonormal basis $\{|x\rangle,x \in {\mathbb{Z}_{N}}= {\mathbb{Z}\,\,(\mbox{mod}{N})}\}.$  The {\em coin space} is the Hilbert space $\mathcal{H}_2$ spanned by an orthonormal basis $\{|r\rangle, |l\rangle\}$. The ``state vector space" is $\mathcal{H} =  \mathcal{H}_N \otimes \mathcal{H}_2$. Thus, a general density operator $\rho$ in $\mathcal{H}$ may be expressed as 
$$\rho=\sum_{x,y \in \mathbb{Z}_{N}}\sum_{i,j \in\{r,l\}}a_{xi,yj}|xi\rangle\langle yj|.$$

In the sequel, we may use the $2N$ by $2N$ matrix $[a_{xi,yj}]$ to represent the density operator $\rho$.

As in \cite{KT032}, the temporal progression of states of a quantum walk on the $N$-cycle is modeled by repeated iterations of a quantum operation of the form   
\begin{eqnarray}
{\bf \Phi}(\rho)=(1-q)U\rho U^{\dagger}+q \sum_{x\in \mathbb{Z}_{N}}\sum_{i \in\{r,l\}}\mathbb{P}_{xi} U\rho U^{\dagger} \mathbb{P}_{xi}^{\dagger}.\label{special-ruod}
\end{eqnarray}
 
In Eq. (\ref{special-ruod}), the unitary operator $U = S(\mathbb{I}\otimes C)$ is defined in terms of the {\em shift operator} $S : \mathcal{H} \rightarrow \mathcal{H}$, which acts on basis states $|xi\rangle$ by the formulas $S(|xr\rangle)=|(x+1)r\rangle$ and $S(|xl\rangle)=|(x-1)l\rangle$. As always, $\mathbb{I}$ denotes the identity operator, acting, in this case, on $\mathcal{H}_{N}$. Meanwhile, any unitary operator $C:\mathcal{H}_2 \rightarrow \mathcal{H}_2$ is eligible to serve as the {\em coin operator}. In general, $C$ may be expressed by a formula such as $C=u_{11}|r\rangle\langle r|+u_{21}|l\rangle\langle r|+u_{12}|r\rangle\langle l|+u_{22}|l\rangle\langle l|$. However, for technical reasons, it will be convenient, in the sequel, to assume that the complex coefficients $u_{11}$, $u_{12}$, $u_{21}$ and $u_{22}$ all are non-zero. 

In Eq. (9), the parameter $q$ ($0\le q\le 1$) is called the decoherence rate. When $q=0$, the QW evolves as a purely coherent quantum process. At the other extreme, when $q=1$, the QW behaves exactly like a classical random walk. Since, in this paper, we are interested in the case of non-classical quantum walks subject to a non-zero level of decoherence, it shall be assumed henceforth that $0<q<1$. Finally, in Eq. (\ref{special-ruod}), the {\em projection operator}  $\mathbb{P}_{xi}=|xi\rangle\langle xi|$ acts with probability $q$ on the Hilbert space spanned by the eigenstate $|xi\rangle$. Thus, at each time step of the quantum walk generated by the quantum operation ${\bf \Phi}$, exposure to decoherence prevails with probability $q$. 

By comparison with Eq. (\ref{GRO}), we see that the definition of ${\bf \Phi}$ in Eq. (\ref{special-ruod}) conforms to the definition of a generalized random unitary operation (GRO). Thus, ${\bf \Phi}$ falls within the purview of Theorem 1.


For an eigenvalue $\lambda$ of ${\bf \Phi}$, let $E_{\bf \Phi}(\lambda)=\mathsf{Ker}({\bf \Phi}-\lambda \mathbb{I})$ denote the eigenspace of $\lambda$. For ${\bf \Phi}$ as in Eq. (\ref{special-ruod}), the following lemma elucidates the relationship between the various eigenspaces of ${\bf \Phi}$.

We digress momentarily to recall what is meant by the term ``generalized eigenvector". Suppose $J$ is a Jordan block in the Jordan canonical form of ${\bf \Phi}$ corresponding to an eigenvalue $\lambda$ and suppose its dimension is $\mathrm{dim}(J)=m>1$. Let $Y_{1}$ denote the corresponding eigenvector. Then, by standard linear algebra, there exists a sequence  $Y_{1}$, $Y_{2}$, ..., $Y_{m}$ of what are called ``generalized eigenvectors" characterized by the conditions:
\begin{eqnarray}
({\bf \Phi}-\lambda \mathbb{I})Y_{k}=Y_{k-1},
\end{eqnarray}
where $k=1,..., m$ and where, by definition, $Y_{0}=0$. The generalized eigenvectors belonging to $\lambda$ are linearly independent and satisfy the condition $({\bf \Phi}-\lambda \mathbb{I})^kY_{k}=0$.

{\bf Lemma~3.}\, Let $\lambda_1$, $\lambda_2$ be eigenvalues of ${\bf \Phi}$ with $|\lambda_1|=|\lambda_2|=1$. Let $\alpha$ be an eigenvalue of ${\bf \Phi}$ with $|\alpha|<1$ and let $Y_{1}$, ..., $Y_{j_{\alpha}}$ denote the generalized eigenvectors belonging to $\alpha$. Then 
\begin{enumerate}
\item If $\lambda_1\ne \lambda_2$, then $E_{\bf \Phi}(\lambda_1)\perp E_{\bf \Phi}(\lambda_2)$.
\item $E_{\bf \Phi}(\lambda_1)\perp \mathsf{Span}\{Y_{1}, ..., Y_{j_{\alpha}}\}$
\end{enumerate}
Proof. See Appendix B.

For a QW on the $N$-cycle generated by ${\bf \Phi}$ as in Eq. (\ref{special-ruod}), it turns out that the eigenvalues of ${\bf \Phi}$ on the unit circle and the structure of their eigenspaces are completely determined by the parity of the cycle length $N$. The following lemma articulates the details. 

{\bf Lemma 4.}\,\, For a quantum walk on the $N$-cycle governed by ${\bf \Phi}$ as in Eq. (\ref{special-ruod}): 
\begin{enumerate}
\item If $N$ is odd, then 1 is the only eigenvalue on the unit circle, and its eigenspace $E_{\bf \Phi}(1)=\mathrm{span}\{\mathbb{I}_{2N}\}$.
\item If $N$ is even, then 1 and -1 are the only eigenvalues of $\Phi$ on the unit circle, in which case $E_{\bf \Phi}(1)=\mathrm{span}\{\mathbb{I}_{2N}\}$, while $E_{\bf \Phi}(-1)=\mathrm{span}\{\mathbb{I}_{\pm 1}\}$, where $\mathbb{I}_{\pm 1}=\mbox{diag}(1,1,-1,-1\ldots,1,-1,-1)$.
\end{enumerate}
Proof. See Appendix C.

\vskip 0.1in
At this point we have gathered all of the necessary ingredients to formulate a description of the long-term behavior of a quantum walk on the $N$-cycle generated by the quantum operation ${\bf \Phi}$ as given by Eq.(\ref{special-ruod}).

{Theorem 5.}\,\, Suppose a quantum walk, generated by the quantum operation ${\bf \Phi}$ as defined by Eq.(\ref{special-ruod}), is launched on the $N$-cycle with initial state $\rho(0)$ and with decoherence rate $0<q<1$. If $N$ is odd, then the iterated succession of quantum states $\rho(t)={\bf \Phi}^t\rho(0)$ converges to $\frac{1}{2N}\mathbb{I}_{2N}$. If $N$ is even, then $\|{\bf \Phi}^t\rho(0)-\frac{1}{2N}\mathbb{I}_{2N}-(-1)^t\frac{1}{2N}\langle \rho(0),\mathbb{I}_{\pm 1} \rangle \mathbb{I}_{\pm 1}\| $ converges to zero. In particular, if $\langle \rho(0),\mathbb{I}_{\pm 1} \rangle=0$, then $\rho(t)={\bf \Phi}^t\rho(0)$ converges to $\frac{1}{2N}\mathbb{I}_{2N}$.\\
Proof. See Appendix D.

This theorem represents a significant advance compared to our previous work in \cite{LP092}. In that paper, which dealt with a specific model of quantum walks on the $N$-cycle, subject to decoherence on the coin degree of freedom, an explicit formula was derived for the limiting probability distribution, not just in the weak time-averaged sense, but in the strong point-wise sense (see Theorem 3 in \cite{LP092}). As seen below, this result follows as an immediate corollary of Theorem 5 (see Corollary 6 below). However, unlike Theorem 5 of the present paper, Theorem 3 in \cite{LP092} fails to specify the limiting structure of the density matrices themselves. We speculate that the analytical approach employed in this paper might prove fruitful also in the context of \cite{LP092} to specify the limiting structure of the corresponding density matrices.

From this theorem we can derive an immediate corollary concerning the position probability distribution for quantum walks on the $N$-cycle.
Let $P(x,t)=\mathrm{Tr}\left(|x\rangle\langle x|\rho(t)\right)$ denote the probability of finding the walker at the position $x$ at time $t$.

{\bf Corollary 6.}\,\, For a quantum walk launched on the $N$-cycle with decoherence rate $0<q<1$  and driven by the quantum operation ${\bf \Phi}$ as in Eq. (\ref{special-ruod}), the following assertions hold:

(i)\, If $N$ is odd, then $P(x,t)$ converges to $\frac{1}{N}$ on all nodes of the cycle regardless of the initial state $\rho(0)$.

(ii)\, If $N$ is even and if the quantum walk is launched from a definite initial node (i.e., $|\langle \rho(0),\mathbb{I}_{\pm 1} \rangle|=1$) , then $P(x,t)$ converges to $\frac{2}{N}$ on the supporting nodes of the cycle and to 0 on the non-supporting nodes of the cycle.

(iii)\, If $N$ is even and if the parity of the initial node of the QW has an equal probability of being odd or even (i.e., $\langle \rho(0),\mathbb{I}_{\pm 1} \rangle=0$), then $P(x,t)$ converges to $\frac{1}{N}$ on all nodes of the cycle.

Items (i) and (ii) of Corollary 6 confirm the predictions of \cite{KT032, MK07}, which are based on numerical simulations. However, since the class of quantum walks covered by Corollary 6 includes not only those launched in a pure state, as considered by \cite{KT032, MK07}, but also those launched in a mixed state, Corollary 6 actually is a stronger version of the predictions in \cite{KT032, MK07}. 

Our analysis of the long-term evolution of quantum walks on the $N$-cycle generated by a generalized random unitary operation would not be complete without a consideration of the effects of decoherence upon entanglement. We begin by reviewing some preliminaries. 

In what follows, we utilize the concept of von Neumann entropy to quantify the mutual information between subsystems (coin and position). Intuitively, the von Neumann entropy of a quantum system $A$, denoted $S(A)$, is a measure of the uncertainty implied by the multitude of potential outcomes as reflected by its density matrix $\rho(A)$. More precisely, $S(A) = S(\rho(A))=-\mathrm{Tr}(\rho \ln \rho)$. 

For a composite system with two components $A$ and $B$, the joint entropy of their conjunction, denoted by $S(A,B)$, is defined by the formula $S(A,B)=-\mathrm{Tr}(\rho^{AB}\ln \rho^{AB})$, where $\rho^{AB}$ is the density matrix of the composite quantum system $AB$. 

A good measure of the level of quantum entanglement between the two components $A$ and $B$ is the so-called mutual information $S(A:B)$, defined by the formula $S(A:B)=S(A)+S(B)-S(A,B)$.

The following Lemma, due to Watrous \cite{JW08}, is essential to our reasoning. 

{\bf Lemma 7 }\, Let $\mathcal{X}$ denote a complex Euclidean space and let $\mathrm{Pos}(\mathcal{X})$ denote the set of  positive semidefinite operators defined on $\mathcal{X}$ with norm defined by Eq. (\ref{norm}). Then, with respect to this norm, the von Neumann entropy $S(\rho)$ is continuous at every point $\rho \in \mathrm{Pos}(\mathcal{X})$.

For quantum walks on the $N$-cycle, the following definitions apply. For the subsystem associated with the coin, the time-dependent {\it reduced} density operator $\rho_{\mbox{c}}(t)$ is given by $\rho_{\mbox{c}}(t)=\mathrm{trace}_{\mbox{w}}(\rho(t))$, where the subscript \mbox{w} signifies exclusion or ``tracing out", relative to the overall system density operator $\rho(t)$, of the walker's degrees of freedom. Similarly, for the subsystem associated with the walker, the time-dependent {\it reduced} density operator $\rho_{\mbox{w}}(t)$ is given by $\rho_{\mbox{w}}(t)=\mathrm{trace}_{\mbox{c}}(\rho(t))$, where the subscript \mbox{c} signifies exclusion or ``tracing out", relative to the overall system density operator $\rho(t)$, of the degrees of freedom of the coin.

The following theorem summarizes our main findings in connection with the behavior of quantum entanglement for quantum walks on the $N$-cycle.

{\bf Theorem 8.}\, Suppose a quantum walk is launched on the $N$-cycle with initial state $\rho(0)$ and with decoherence rate $0<q<1$,  and driven by the quantum operation ${\bf \Phi}$ as in Eq. (\ref{special-ruod}). Let $\rho_{\mbox{c}}(t)$ and $\rho_{\mbox{w}}(t)$ denote, respectively, the time-dependent reduced density operators associated with the subsystems of the coin and the walker. Then the long-term trend of the mutual information between the coin subsystem and the walker subsystem is given by $\lim_{t\rightarrow \infty}S\left(\rho_{\mbox{c}}(t):\rho_{\mbox{w}}(t)\right)=0$.
 
Proof. See appendix E.

To put it briefly, according to Theorem 8, exposed to any non-zero level of decoherence, the mutual quantum information between subsystems tends to zero, signifying the total collapse of entanglement between subsystems.  

\section{Conclusions}

For the bipartite quantum system considered in this paper, our analysis has shown that exposure to any nonzero level of persistent decoherence causes the system to behave asymptotically like a purely classical system. As noted in \cite{Z03}, decoherence on a quantum system is manifested through its density matrix by the vanishing of the off-diagonal elements. Ultimately, the density operators should become indistinguishable from diagonal matrices. In the context of a coin-driven quantum walk on the $N$-cycle, this is precisely what we assert in Theorem 5. Note that the off-diagonal elements of the density matrix are precisely the elements that represent the quantum correlations (also known as entanglement) between the coin subsystem and the position subsystem. Unsurprisingly, at least for quantum walks, as shown in Theorem 8, decoherence turns out to be practically synonymous with ``disentanglement", so to speak. Indeed, as a corollary of our results, we show, when influenced by decoherence, the resulting long-term distributions are indistinguishable from those exhibited by classical random walks.

The model of decoherence used in this article is only one of several prevalent in the current literature. It would be interesting to investigate how quantum entanglement and probability distribution respond to other models of decoherence (such as amplitude damping, or  phase damping acting on the coin degree of freedom ), and not just for quantum walks on the $N$-cycle, but for quantum walks over other kinds of topological networks as well. For a model of a quantum state transition, the long-term behavior of such associated quantum Markov chains is intimately linked to the structure of the eigenspaces of eigenvalues on the unit circle. We speculate that the characterization of the linear structure of its key eigenspaces in Theorem 1, might provide an efficient means for identifying the eigenspaces of all eigenvalues of absolute value 1, thereby enabling one to explore the asymptotic evolution of the associated Markov chains determined by a generalized random unitary operation in Eq. (\ref{GRO}).

\begin{acknowledgments}

CL was supported by NSF grant CCF-1005564.
 
\end{acknowledgments}

\appendix

\section{PROOF OF THEOREM 1}

Proof. \, If for each index $1\leq i\leq k$ we have $U_iX=\lambda XU_i$ and $U_iXU_i^{\dagger}=\sum_{j=1}^lA_jXA_j^{\dagger}$, then it is easy to see that $X\in \mathrm{Ker}(\Psi-\lambda \mathbb{I})$. We proceed to justify the assertion in the opposite direction. Firstly, by \cite{PGWPR06}, the following inequality must be true:  $\|\sum_{j=1}^l A_j X A_j^{\dagger}\|\le \|X\|$. If $X\in\mathrm{Ker}(\Psi-\lambda \mathbb{I})$, then using the unitary invariance of the Hilbert-Schmidt norm and the contractivity of a completely positive trace-preserving linear map,  we get
\begin{eqnarray}
\|X\|=\|\lambda X\|=\|\sum_{i=1}^k p_iU_iX U_i^{\dagger}+q\sum_{j=1}^l A_j X A_j^{\dagger}\| \nonumber\\
\le \sum_{i=1}^kp_i\| U_i X U_i^{\dagger}\|+q\|\sum_{j=1}^l A_j X A_j^{\dagger}\| \nonumber\\
\le \sum_{i=1}^kp_i\|X\|+q\|X\|=\|X\|.\label{basic-1}
\end{eqnarray}

The inequality (\ref{basic-1}) actually is an equality. In particular, we have
\begin{eqnarray}
\|\sum_{j=1}^l A_j X A_j^{\dagger}\|=\|X\|\label{specialbasic}
\end{eqnarray}

For convenience, we set $p_{k+1}=q$, $v_i=U_iXU_i^{\dagger}$, $i=1, 2, ..., k$; and $v_{k+1}=\sum_{j=1}^l A_j X A_j^{\dagger}$. Then Eq. (\ref{basic-1}) implies the following equality. 
\begin{eqnarray}
\langle \sum_{i=1}^{k+1} p_iv_i,\sum_{i=1}^{k+1} p_iv_i\rangle=(\sum_{i=1}^{k+1}p_i\langle v_i,v_i\rangle^{\frac{1}{2}})^2. \label{basic-2}
\end{eqnarray}
Eq.(\ref{basic-2}) leads to

\begin{eqnarray}
\sum_{i<j}2p_ip_j\langle v_i,v_i\rangle^{\frac{1}{2}}\langle v_j,v_j\rangle^{\frac{1}{2}}\nonumber\\
=\sum_{i<j}p_ip_j[\langle v_i, v_j\rangle+\langle v_j,v_i\rangle]\nonumber\\
=\sum_{i<j}2p_ip_j\mathrm{Re}(\langle v_i,v_j\rangle)\nonumber\\
\le \sum_{i<j}2p_ip_j|\langle v_i,v_j\rangle|\nonumber\\
\le\sum_{i<j}2p_ip_j\langle v_i,v_i\rangle^{\frac{1}{2}}\langle v_j,v_j\rangle^{\frac{1}{2}}.\label{basic-3}
\end{eqnarray}

Because of the identity of both right and left hand sides of the algebraic expression (\ref{basic-3}), all inequalities involved are in fact are equalities. It can be readily seen that 
\begin{eqnarray}
\mathrm{Re}(\langle v_i, v_j\rangle)=|\langle v_i, v_j\rangle|=\langle v_i, v_i\rangle^{\frac{1}{2}}\langle v_j, v_j\rangle^{\frac{1}{2}}\ne 0\nonumber\\
\mathrm{for}\, i, j \in \{1, 2, ..., k+1\}.\label{basic-4}
\end{eqnarray}
According to Cauchy-Schwarz inequality, the second equality of Eq. (\ref{basic-4}) implies that $v_i=\beta_{ij}v_j$ for all possible pair $(i,j)$ with $i<j$. Here $\beta_{ij}$ a complex number. The first equality of Eq. (\ref{basic-4}) further implies every $\beta_{ij}$ must be positive.

Due to the unitary invariance of the Hilbert-Schmidt norm, we have $\beta_{ij}=1$ for all pairs $(i,j)$ where $i<j\le k$. When $j=k+1$, since $\|X\|=\|v_i\|=\beta_{i(k+1)}\|v_{k+1}\|=\beta_{i(k+1)}\|\sum_{j=1}^l A_j X A_j^{\dagger}\|=\beta_{i(k+1)}\|X\|$ based on Eq. (\ref{specialbasic}), this implies that $\beta_{i(k+1)}=1$ for all $i<k+1$.

Overall we have 
\begin{eqnarray}
U_1XU_1^{\dagger}=U_2XU_2^{\dagger}=...=U_kXU_k^{\dagger}=\sum_{j=1}^l A_j X A_j^{\dagger}.
\end{eqnarray}
Noticing that $\Psi(X)=\lambda X$, hence we get $U_iXU_i^{\dagger}=\lambda X$ for $i\in\{1, 2,..., k\}$, this completes the justification.

\section{PROOF OF LEMMA 3}

Proof.\,\, The adjoint operator of $\Phi$ defined in Eq. (\ref{special-ruod}) is given by

\begin{eqnarray}
{\bf \Phi}^{\dagger}(\rho)=(1-q)U^{\dagger}\rho U+q \sum_{x\in \mathbb{Z}_{N}}\sum_{i\in \{r,l\}}U^{\dagger}\mathbb{P}^{\dagger}_{xi}\rho \mathbb{P}_{xi} U. \label{adruod}
\end{eqnarray}

We shall first justify the following assertion:

If ${\bf \Phi}(X)=\lambda X$ where $|\lambda|=1$, then ${\bf \Phi}^{\dagger}(X)=\lambda^{\ast} X$.

Theorem 1 implies that $U^{\dagger}XU=\lambda^{\ast} X$ and  $X=\sum_{x\in \mathbb{Z}_{N}}\sum_{i\in \{r,l\}}\mathbb{P}_{xi} X\mathbb{P}_{xi}^{\dagger}$. Note that ${\bf \Phi}^{\dagger}(X)=(1-q)U^{\dagger}X U+q \sum_{x\in \mathbb{Z}_{N}}\sum_{i\in \{r,l\}}U^{\dagger}\mathbb{P}^{\dagger}_{xi}X\mathbb{P}_{xi} U$, this implies that ${\bf \Phi}^{\dagger}(X)=\lambda^{\ast}X$.

To prove statement 1, let $Z \in E_{\bf \Phi}(\lambda_1)$ and $Y\in E_{\bf \Phi}(\lambda_2)$. Then, by the assertion above,  $\lambda_1\langle Z, Y\rangle=\langle {\bf \Phi}^{\dagger} Z, Y\rangle=\langle Z, {\bf \Phi}Y\rangle=\lambda_2\langle Z,Y\rangle$. Since $\lambda_1 \ne \lambda_2$, it follows that $\langle Z, Y\rangle=0$. Hence $E_{\bf \Phi}(\lambda_1)\perp E_{\bf \Phi}(\lambda_2)$.

We proceed to justify statement 2. For an eigenvalue $\alpha$ with $|\alpha|<1$, we may assume, without loss of generality, that the generalized eigenvectors belonging to $\alpha$ are arranged in a sequence $Y_1$, $Y_2$, ..., $Y_{j_{\alpha}}$ such that:
\begin{eqnarray}
({\bf \Phi}-\alpha\mathbb{I})Y_r=Y_{r-1},
\end{eqnarray}
where, by definition, $Y_0 = 0$. It follows
that $({\bf \Phi}-\alpha\mathbb{I})^r Y_r = 0$. If $Z\in \mathsf{Ker}({\bf \Phi}-\lambda_1 \mathbb{I})$, then, again by the assertion,  $\lambda_1 \langle Z, Y_1\rangle=\langle {\bf \Phi}^{\dagger}Z, Y_1\rangle=\langle Z, {\bf \Phi}Y_1\rangle=\alpha\langle Z, Y_1\rangle$, which implies that $\langle Z, Y_1\rangle=0$. Similarly, $\lambda_1\langle Z, Y_2\rangle=\langle {\bf \Phi}^{\dagger}Z, Y_2\rangle=\langle Z, {\bf \Phi}Y_2\rangle=\langle Z, Y_1\rangle+\alpha\langle Z, Y_2\rangle$, from which it follows that $\langle Z, Y_2\rangle=0$. Likewise, by the same reasoning, applied repeatedly, we deduce that $\langle Z, Y_r\rangle=0$ for any $r$. Thus $E_{\bf \Phi}(\lambda_1)\perp \mathsf{Span}\{Y_1, Y_2, ..., Y_{j_{\alpha}}\}$. 

\section{PROOF OF LEMMA 4}

Proof.\,\, Let $\lambda$ be an eigenvalue of ${\bf \Phi}$ with $|\lambda|=1$. Picking $X \in \mathsf{Ker}({\bf \Phi}-\lambda \mathbb{I})$ such that $X\ne 0$, according to Theorem 1,  we have $\sum_{x\in \mathbb{Z}_{N}}\sum_{i\in \{r,l\}}\mathbb{P}_{xi} UXU^{\dagger} \mathbb{P}_{xi}^{\dagger}=UXU^{\dagger}$ which implies that $UXU^{\dagger}$ is a diagonal matrix. Since $UXU^{\dagger}=\lambda X$, then $X$ must be diagonal. Without loss of generality, it may be assumed that $X=\sum_{x\in \mathbb{Z}_N}a_{xr}|xr\rangle\langle xr|+\sum_{x\in \mathbb{Z}_N}a_{xl}|xl\rangle\langle xl|$. Note that $U$, as a linear operator, can be expressed by 
\begin{eqnarray}
U=\sum_{x\in \mathbb{Z}_N}[u_{11}|(x+1)r\rangle\langle xr|+u_{21}|(x-1)l\rangle\langle xr|\nonumber\\
+u_{12}|(x+1)r\rangle\langle xl|+u_{22}|(x-1)l\rangle\langle xl|].
\end{eqnarray}
Therefore it can readily deduced that
\begin{eqnarray}
\!\!\!UX
\!\!&=&\!\!\sum_{x\in \mathbb{Z}_N}a_{xr}u_{11}|(x+1)r\rangle\langle xr| \nonumber \\
\!\!&+&\sum_{x\in \mathbb{Z}_N}a_{xr}u_{21}|(x-1)l\rangle\langle xr|    \nonumber\\
\!\!&+&\sum_{x\in \mathbb{Z}_N}a_{xl}u_{12}|(x+1)r\rangle\langle xl| \nonumber\\
\!\!&+&\sum_{x \in \mathbb{Z}_N}a_{xl}u_{22}|(x-1)l\rangle\langle xl|;\label{ux}
\end{eqnarray}

\begin{eqnarray}
\!\!\!XU
\!\!&=&\sum_{x\in \mathbb{Z}_N}a_{(x+1)r}u_{11}|(x+1)r\rangle\langle xr| \nonumber\\
\!\!&+&\sum_{x\in \mathbb{Z}_N}a_{(x-1)l}u_{21}|(x-1)l\rangle\langle xr| \nonumber\\
\!\!&+&\sum_{x \in \mathbb{Z}_N}a_{(x+1)r}u_{12}|(x+1)r\rangle\langle xl| \nonumber\\
\!\!&+&\sum_{x \in \mathbb{Z}_N}a_{(x-1)l}u_{22}|(x-1)l\rangle\langle xl|.\label{xu}
\end{eqnarray}


Based on $UX=\lambda XU$, we derive the following four identities.
\begin{eqnarray}
a_{xr}u_{11}=\lambda a_{(x+1)r}u_{11},\quad a_{xr}u_{21}=\lambda a_{(x-1)l}u_{21}.\nonumber\\
a_{xl}u_{12}=\lambda a_{(x+1)r}u_{12},\quad a_{xl}u_{22}=\lambda a_{(x-1)l}u_{22}.\label{identity-1}
\end{eqnarray}
Here none of $u_{11}$,$u_{12}$,$u_{21}$ and $u_{22}$ is zero, and $x\in\mathbb{Z}_{N}$. A little algebraic manipulation of these four identities implies that $\lambda^2=1$ and $\lambda^N=1$. We arrive at the following two cases depending on the parity of $N$.

(1)\, If $N$ is odd, then $\lambda=1$. By equations in (\ref{identity-1}), it is seen that $X=k\mathbb{I}_{2N}$ where $k$ is constant, and so $E_{\bf \Phi}(1)=\mathrm{span}\{\mathbb{I}_{2N}\}$.

(2)\, If $N$ is even, then $\lambda=1$ or $\lambda=-1$. In the case when $\lambda=1$, $X=k\mathbb{I}_{2N}$ where $k$ is constant, and so $E_{\bf \Phi}(1)=\mathrm{span}\{\mathbb{I}_{2N}\}$. In the case when $\lambda=-1$, $X=k\mathbb{I}_{\pm 1}$ where $k$ is constant, and $E_{\bf \Phi}(-1)=\mathrm{span}\{\mathbb{I}_{\pm 1}\}$.

\section{PROOF OF THEOREM 5}

Proof.\, In the case when $N$ is odd, $\lambda$= 1 is the only eigenvalue of ${\bf \Phi}$ on the unit circle by Lemma 4, therefore the conclusion follows upon applying theorem 5 in \cite{LPQMC10}.

In the case of $N$ is even, by Lemma 4, we have $E_{{\bf \Phi}}(1)=\mathrm{Span}\{\mathbb{I}_{2N}\}$ and $E_{{\bf \Phi}}(-1)=\mathrm{Span}\{\mathbb{I}_{\pm 1}\}$. In this context, we may choose for $\mathfrak{B}(\mathcal{H})$ a basis consisting of a basis $Z_1=\frac{1}{\sqrt{2N}}\mathbb{I}_{2N}$ for $E_{{\bf \Phi}}(1)$, a basis $Z_{-1}=\frac{1}{\sqrt{2N}}\mathbb{I}_{\pm 1}$ for $E_{{\bf \Phi}}(-1)$, and together with an orthogonally complement basis consisting of generalized eigenvectors belonging to all other eigenvalues. In terms of such a basis we have $\rho(0)=c_1 Z_1\oplus c_{-1}Z_{-1} \oplus W$ where $W \perp Z_{r}$ for all $r=-1, 1$ by Lemma 3.  By simple linear algebra, it follows that $c_r=\mathrm{tr}( Z_r^{\dagger}\rho(0))$ for $r=-1, 1$.  

Relative to this basis, the Jordan canonical matrix representation of ${\bf \Phi}$ is given by:
\begin{eqnarray}
\left[{\bf \Phi}\right]=\mbox{diag}\left(1, -1, J_1, J_2, ...,J_h\right), 
\end{eqnarray}
where $J_r$ is the Jordan block corresponding to the eigenvalue $\alpha_r$ of the magnitude strictly less than 1, $r=1, 2, ..., h$.

Consider what becomes of the Jordan blocks of the powers $\left[{\bf \Phi}\right]^t$ as $t\rightarrow \infty$. Since each of the Jordan blocks $J_r$ is an upper triangular matrix whose diagonal is populated by a single eigenvalue of modulus strictly less than unity, it is a simple exercise in elementary algebra to show that $\lim_{t\rightarrow\infty}J^{t}_{r}=O_{r}$ (zero matrix of same size as $J_{r}$). Thus, if we define 
\begin{eqnarray}
\left[{\bf \Phi}^{\infty}\right](t)=\mbox{diag}\left(1, (-1)^t, O_1, O_2, ...,O_h\right), \label{asymptotic-form}
\end{eqnarray}
then $\|{\bf \Phi}^{t}-{\bf \Phi}^{\infty}(t)\|\rightarrow 0$.

Therefore $\|{\bf \Phi}^t\rho(0)-\frac{1}{2N}\mathbb{I}_{2N}-(-1)^t\frac{1}{2N}\langle \rho(0),\mathbb{I}_{\pm 1} \rangle \mathbb{I}_{\pm 1}\| $ converges to zero; In particular, if $\langle \rho(0),\mathbb{I}_{\pm 1} \rangle=0$, $\rho(t)={\bf \Phi}^t\rho(0)$ converges to $\frac{1}{2N}\mathbb{I}_{2N}$. 

\section{PROOF OF THEOREM 8}

Proof.\, We assume that $N$ is odd. By Theorem 5, $\lim_{t\rightarrow \infty}\rho(t)=\frac{1}{2N}\mathbb{I}_{2N}$, which implies that $\lim_{t\rightarrow \infty}S(\rho(t))=S(\frac{1}{2N}\mathbb{I}_{2N})$ by Lemma 7. Since both $\mathrm{trace}_{\mbox{w}}\left(\cdot\right)$ and $\mathrm{trace}_{\mbox{c}}\left(\cdot\right)$ are continuous functions of the argument, it follows, by Theorem 5, that $\lim_{t\rightarrow \infty}\rho_{\mbox{c}}(t)=\mathrm{trace}_{\mbox{w}}\left(\frac{1}{2N}\mathbb{I}_{2N}\right)$ and $\lim_{t\rightarrow \infty}\rho_{\mbox{w}}(t)=\mathrm{trace}_{\mbox{c}}\left(\frac{1}{2N}\mathbb{I}_{2N}\right)$. These imply that $\lim_{t\rightarrow \infty}S(\rho_{\mbox{c}}(t))=S(\mathrm{trace}_{\mbox{w}}\left(\frac{1}{2N}\mathbb{I}_{2N}\right))$ and $\lim_{t\rightarrow \infty}S(\rho_{\mbox{w}}(t))=S(\mathrm{trace}_{\mbox{c}}\left(\frac{1}{2N}\mathbb{I}_{2N}\right))$ by Lemma 7. Because of $S(\frac{1}{2N}\mathbb{I}_{2N})=S(\mathrm{trace}_{\mbox{w}}\left(\frac{1}{2N}\mathbb{I}_{2N}\right))+S(\mathrm{trace}_{\mbox{c}}\left(\frac{1}{2N}\mathbb{I}_{2N}\right))$, so $\lim_{t\rightarrow \infty}S\left(\rho_{\mbox{c}}(t):\rho_{\mbox{w}}(t)\right)=0$.

In the case when $N$ is even, we consider two sub-cases: (a) $t$ is odd; and (b) $t$ is even. In each of the two sub-cases, the assertion is proved in exactly the same way as in the case when $N$ is odd.

\bibliography{apssamp}

\smallskip

\begin{thebibliography}{9}



\bibitem{CG04}

A. M. Childs and J. Goldstone, Phys. Rev. A 70, 022314 (2004).

\bibitem{A07}

A. Ambainis, SIAM J. Comput. 37, pp. 210-239 (2007).  

\bibitem{PGKJ09}

V. Poto$\check{\mathrm{c}}$ek, A. G$\acute{\mathrm{a}}$bris, T. Kiss, and I. Jex, Phys. Rev. A 79, 012325 (2009). 


\bibitem{VA08}

S.E. Venegas-Andraca, {\em Quantum Walks for Computer Scientists}, Morgan and Claypool Publishers (Synthesis Lectures on Quantum Computing), 2008.

\bibitem{K03}

J. Kempe, Contemp. Phys. {\bf 44}, 307 (2003).



\bibitem{K08}

N. Konno, in {\em Quantum Potential Theory}, Lecture Notes in Mathematics, edited by U. Franz and M. Schurmann (Springer-Verlag, Heidelberg, 2008), pp.309-452.

\bibitem{GJS04}

G. Grimmett, S. Janson and P. F. Scudo, Phys. Rev. E {\bf 69}, 026119 (2004).

\bibitem{SM10}

H. Schmitz, R. Matjeschk, Ch. Schneider, J. Glueckert, M. Enderlein, T. Huber, and T. Schaetz, Phys. Rev. Lett. 103, 090504 (2009).

\bibitem{ZKG10}

F. Z$\ddot{\mathrm{a}}$hringer, G. Kirchmair, R. Gerritsma, E. Solano, R. Blatt, and C. F. Roos, Phys. Rev. Lett. 104, 100503 (2010).

\bibitem{KFC09}

M. Karski, L. F$\ddot{\mathrm{o}}$ster, J.-M. Choi, A. Steffen, W. Alt, D. Meschede, and A. Widera, Science 325, 174 (2009).

\bibitem{ZLLLLG10}

 P. Zhang, B. H. Liu, R. F. Liu, H. R. Li, F. L. Li, and G. C. Guo, Phys. Rev A {\bf 81}, 052322 (2010).

\bibitem{SCP10}

A. Schreiber, K. N. Cassemiro, V. Poto$\check{\mathrm{c}}$ek, A. G$\acute{a}$bris, P. J. Mosley, E. Andersson, I. Jex, and Ch. Silberhorn, Phys. Rev. Lett. 104, 050502 (2010).

\bibitem{BFL10}
M. A. Broome, A. Fedrizzi, B. P. Lanyon, I. Kassal, A. Aspuru-Guzik, and A. G. White, Phys. Rev. Lett. 104, 153602 (2010).


\bibitem{BCA022} 

T. A. Brun,, H. A. Carteret and A. Ambainis,  Phys. Rev. A 67, 032304 (2003).


\bibitem{KT032}
V. Kendon and B. Tregenna, Phys. Rev. A 67, 042315 (2003). 

\bibitem{RSAAD05}
A. Romanelli, R. Siri, G. Abal, A. Auyuanet and R. Donangelo, Physica A 347, 137-152(2005).

\bibitem{R07}

P. C. Richter, Phys. Rev. A 76, 042306(2007).


\bibitem{BSCR08}

S. Banerjee, R. Srikanth, C. M. Chandrashekar and P. Rungta, Phys. Rev. A 78, 052316 (2008).

\bibitem{LP092}

C. Liu and N. Petulante, Phys. Rev. E 81, 031113(2010).

\bibitem{AAAD09}

M. Annabestani, S. J. Akhtarshenas and M. R. Abolhassani, Phys. Rev. A 81, 032321 (2010).

\bibitem{SCSK10}

Y. Shikano, K. Chisaki, E. Segawa and N. Konno, Phys. Rev. A 81 062129 (2010).

\bibitem{K06}

 V. Kendon, Math. Struct. Comp. Sci 17, 1169(2006).

\bibitem{AAKV01}

D. Aharanov, A. Ambainis, J. Kempe and U. Vazirani, in Proceedings of the 33rd Annual ACM Symposium on Theory of Computing, (ACM, New York, 2001), pp.50-59.

\bibitem{TFMK03}

B. Tregenna, W. Flanagan, R. Maile, and V. Kendon, New J. Phys. {\bf 5}, 83 (2003).

\bibitem{BGKLW03}

M. Bednarska, A. Grudka, P. Kurzy$\acute{\mathrm{n}}$ski, T. $\L$uczak, and  A. W$\acute{\mathrm{o}}$jcik, Phys. Lett. A 317 Issues 1-2, pp.21–-25(2003). 


\bibitem{IKKS05}
N. Inui, Y. Konishi, N. Konno and T. Soshi, International Journal of Quantum Information, Vol.3, No.3, 535-550 (2005).


\bibitem{LP10}

C. Liu and N. Petulante (2010),  Math. Struct. in Comp. Science, vol. 20, pp. 1099-1115.







\bibitem{MK07}

O. Maloyer and  V. Kendon, New J. Phys. {\bf 9} 87 (2007).

\bibitem{ASH01}

A. S. Holevo, {\it Statistical Structure of Quantum Theory} (Springer, Berlin, 2001).

\bibitem{AS08}

K. Audenaert and S. Scheel, New J. Phys. {\bf 10} 023011(2008).

\bibitem{NAJ10}

J. Novotn$\acute{\mathrm{y}}$, G. Alber and I. Jex, Cent. Eur. J. Phys. DOI: 10.2478/s11534-010-0018-8.


\bibitem{NC00}

M. A. Nielsen and I. L. Chuang, Quantum Computation and Quantum Information, (Cambridge University Press, Cam-
bridge, 2000).

\bibitem{P08}

D. Petz, Quantum Information Theory and Quantum Statistics, Theoretical and Mathematical Physics (Springer, Berlin Heidelberg 2008).

\bibitem{Choi1975}

M.D. Choi, Lin. Alg. Appl. {\bf 10}, 285-290(1975).

\bibitem{K1983}

K. Kraus (1983), States, Effects and Operations: Fundamental Notions of Quantum Theory, Springer Verlag, 1983.

\bibitem{NAJ09}

J. Novotny, G. Alber and I. Jex, J. Phys. A: Math. Theor. {\bf 42}, 282003(2009).

\bibitem{PGWPR06}

D.P$\acute{{\mathrm e}}$rez-Garc$\acute{\mathrm{i}}$a, M. M. Wolf, D. Petz, M. B. Ruskai, J. Math. Phys. {\bf 47}, 083506 (2006).

\bibitem{MW10}

M. M. Wolf (2010), {\it Quantum channels $\&$ operations guided tour}, online lecture notes http://www.nbi.dk/~wolf/notes.pdf, September 12, 2010.

\bibitem{LPQMC10}

C. Liu and N. Petulante, On limiting distributions of quantum Markov chains, arXiv:1010.0741 (2010), to appear in International Journal of Mathematics and Mathematical Sciences.


\bibitem{JW08}

J. Watrous,  Theory of Quantum Information, Lecture notes from Fall 2008, Institute for Quantum Computing, University of Waterloo, Canada.

\bibitem{Z03}

W.\, H. Zurek, Decoherence and the transition from quantum to classical -- REVISITED. Decoherence Poincar$\acute{\mathrm{e}}$ Seminar 2005 , Progress in Mathematical Physics, edited by Bertrand Duplantier, Jean-Michel Raimond and Vincent Rivasseau (Birkhäuser Verlag Basel, 2006 ), pp. 1-31. 


\end{thebibliography}

\end{document}